# The life cycle of *Drosophila* orphan genes

Nicola Palmieri[1], Carolin Kosiol[1], Christian Schlötterer[1*]

[1] Institut für Populationsgenetik, Vetmeduni Vienna, Veterinärplatz 1, 1210 Wien, Austria/Europe

Email addresses:

    NP: nicola.palmieri@vetmeduni.ac.at

    CK: carolin.kosiol@vetmeduni.ac.at

    CS: christian.schloetterer@vetmeduni.ac.at

[*]Corresponding author

## Abstract

Orphans are genes restricted to a single phylogenetic lineage and emerge at high rates. While this predicts an accumulation of genes, the gene number has remained remarkably constant through evolution. This paradox has not yet been resolved. Because orphan genes have been mainly analyzed over long evolutionary time scales, orphan loss has remained unexplored. Here we study the patterns of orphan turnover among close relatives in the *Drosophila obscura* group. We show that orphans are not only emerging at a high rate, but that they are also rapidly lost. Interestingly, recently emerged orphans are more likely to be lost than older ones. Furthermore, highly expressed orphans with a strong male-bias are more likely to be retained. Since both lost and retained orphans show similar evolutionary signatures of functional conservation, we propose that orphan loss is not driven by high rates of sequence evolution, but reflects lineage specific functional requirements.

# Introduction

Orphans are genes with limited phylogenetic distribution and represent a considerable fraction (up to 30%) of the gene catalog in all sequenced genomes (Khalturin, Hemmrich et al. 2009). Studies conducted in different eukaryotes showed that orphans emerge at high rates (Domazet-Loso, Brajkovic et al. 2007; Wissler, Gadau et al. 2013). While gene duplication and exaptation from transposable elements often result in orphan genes (Toll-Riera, Bosch et al. 2009), they also originate frequently *de novo* from non-coding DNA (Cai, Zhao et al. 2008; Heinen, Staubach et al. 2009; Knowles and McLysaght 2009; Wu, Irwin et al. 2011; Yang and Huang 2011; Xie, Zhang et al. 2012; Neme and Tautz 2013; Wu and Zhang 2013), probably through intermediate proto-genes (Carvunis, Rolland et al. 2012). Compared to evolutionary conserved genes, orphans are overall shorter (Lipman, Souvorov et al. 2002), fast evolving (Domazet-Loso and Tautz 2003), have lower and more tissue-restricted expression (Lemos, Bettencourt et al. 2005). Moreover, they often show testis-biased expression (Levine, Jones et al. 2006; Begun, Lindfors et al. 2007), probably due to frequent origination in testis (Kaessmann 2010).

In *Drosophila* the rate of orphan emergence is particulary high (Domazet-Loso, Brajkovic et al. 2007) and many orphans become quickly essential (Chen, Zhang et al. 2010). Although the function of only a few orphan genes has been studied, it has been proposed that orphans might serve an important role in speciation and adaptation to different environments (Khalturin, Anton-Erxleben et al. 2008; Khalturin, Hemmrich et al. 2009; Colbourne, Pfrender et al. 2011). The high rate of orphan origination would predict an increase in gene content over time. However, gene content in eukaryotes spans is remarkably stable compared to genome size, as highlighted by (Tautz and Domazet-Loso 2011). To solve this paradox Tautz and Domazet-Loso proposed that orphans have only a short lifetime ("rapid-turnover" hypothesis) (Tautz and Domazet-Loso 2011). Thus, although orphans are continuously created, most of them might be lost in a relatively short evolutionary time. Relaxed selective costraints in orphans (Cai and Petrov 2010) might also contribute to the high rate of orphan loss.

Moreover, since orphans are typically identifed by the comparison of distantly related species, their evolutionary stability has been so far neglected. This contrasts the comprehensive analysis of evolutionary patterns of gains and losses of non-orphan genes (Hahn, Han et al. 2007). Here, several partially interrelated factors affect gene loss, including gene expression levels, number of protein-protein interactions, gene dispensability and rate of sequence evolution (Krylov, Wolf et al. 2003; Cai and Petrov 2010).

This study focuses on the evolutionary stability of orphan genes. We investigate the factors contributing to orphan loss and find that orphan age, male-biased gene expression and microsatellite content are correlated with orphan stability. Surprisingly, differences in evolutionary rates cannot explain orphan loss and we propose that orphan loss is driven by lineage-specific evolutionary constraints. Overall, orphan genes are lost at a significantly higher rate than non-orphan genes, supporting the "rapid-turnover" hypothesis.

# Results

Orphans are commonly detected by BLASTing the genes of a given organism against a set of outgroup species (Domazet-Loso and Tautz 2003; Toll-Riera, Bosch et al. 2009). A BLASTP cutoff of $10^{-3}$-$10^{-4}$ was found to be optimal to maximize sensitivity and specificity in *Drosophila* (Domazet-Loso and Tautz 2003). To identify orphans we used a BLASTP cutoff of $10^{-4}$ combined with a TBLASTN cutoff of $10^{-4}$, to exclude genes with unannotated orthologs in other species. Following these criteria, we searched in *D. pseudoobscura* for genes with no sequence conservation in 10 *Drosophila* species outside the *D. obscura* group (Supplementary File 1). In total, we identified 1152 orphans, corresponding to 7% of all the *D. pseudoobscura* genes. Our estimate is slightly lower than a previous one (Zhang, Vibranovski et al. 2010), due to our different filtering procedure, but still consistent with a high rate of orphan gain in *Drosophila* (Domazet-Loso and Tautz 2003; Domazet-Loso, Brajkovic et al. 2007; Zhou, Zhang et al. 2008; Wissler, Gadau et al. 2013). Our data clearly indicate that orphan genes are subject to purifying selection, as they show several hallmarks of functional protein coding sequences (Figure 1, 2). A comparison of orphan genes preserved between *D. pseudoobscura* and *D. affinis* resulted in a distribution of *dN/dS* of significantly lower than 1 with a median of 0.44 (Figure 1 – figure supplement 1, one-sided Wilcoxon signed-rank test, $P < 1.0 \times 10^{-15}$), as expected for protein coding sequences. Moreover, *dN/dS* for orphans is significantly lower (Mann-Whitney test, $P = 2.7 \times 10^{-14}$) than *dN/dS* calculated on a random set of intergenic regions with the same length distribution of orphans (see Methods, section "Evolutionary rates") (Figure 1A). Consistent with this, we also found orphans to be more conserved than intergenic regions (Figure 1B, Figure 1 – figure supplement 2.

The codon usage bias of orphans is intermediate to that of old genes and intergenic regions (Figure 1C).

To further test for purifying selection acting on orphans, we used a polymorphism dataset of 45 strains being re-sequenced for the third chromosome of *D. pseudoobscura* (see Methods). We calculated the ratio of synonymous to non-synonymous polymorphism (*pN/pS*), since it provides an indication of purifying selection. We found that *pN/pS* for orphans is significantly lower compared to intergenic regions (Mann-Whitney test, P = 0.02182) (Figure 2), and significantly greater than for old genes (Mann-Whitney test, $P < 1.0 \times 10^{-15}$).

In agreement with studies in other species (Domazet-Loso and Tautz 2003; Toll-Riera, Bosch et al. 2009; Wolf, Novichkov et al. 2009; Cai and Petrov 2010; Capra, Pollard et al. 2010; Carvunis, Rolland et al. 2012) we also find that orphan genes are shorter (median length for orphans = 344 bp, median length for old genes = 1470 bp), have a lower GC content (median GC content for orphans = 0.54, median GC content for old genes = 0.55), are expressed at lower levels (expression in *D. pseudoobscura* males: mean expression for orphans = 29 FPKM, mean expression for old genes = 41 FPKM) than old genes (Figure 3). Using CD-Hit (Li and Godzik 2006) we found the fraction of genes with a paralog (>90% protein similarity) to be similar for orphans (6.9%) and old genes (6.4%). Orphans are more enriched in microsatellites, also consistent with previous findings in vertebrates (Toll-Riera, Rado-Trilla et al. 2012) and rice (Guo, Li et al. 2007). Furthermore, unlike mammals (Toll-Riera, Castelo et al. 2009), none of the *D. pseudoobscura* orphans was found to be associated with transposable elements (see Methods, section "Transposon detection").

The distribution of orphans is heterogeneous across chromosomes ($\chi^2$-test, P < 1.0 × $10^{-15}$), with the X chromosome having the highest fraction of orphans. In the *obscura* group, the two X-chromosome arms have a different evolutionary history. XL corresponds to Muller's element A and is homologous to the X chromosome in *D. melanogaster*. XR, however, has been recently derived from an autosome (Muller's element D, 3L in *D. melanogaster*). Analyzing the old-X and neo-X chromosomes separately, we observed a striking difference in the number of orphans despite similar chromosome sizes, with the old-X responsible for the excess of X-linked orphan genes, and the neo-X showing a similar number of orphans as the autosomes (Figure 4). For each chromosomal arm we computed genomic features in 100 kb windows to correlate them with the difference in orphan content between old-X and neo-X. We found that average GC content, microsatellite density, transposon density and length of intergenic regions differs between the two chromosomal arms (Figure 5).

We hypothesized that this pronounced difference between the two chromosome arms might reflect a different history of X-linkage. If orphan genes emerge at a higher rate on the X-chromosome (Levine, Jones et al. 2006), the shorter history of X-linkage on the neo-X could explain the paucity of orphans on the neo-X compared to old-X. In this case the difference in orphan number between old-X and neo-X chromosomes should date back to the time before the origin of the neo-X, with a similar number of orphans originating after the creation of the neo-X. We therefore used the genomic sequences of five members of the *D. obscura* group (*D. pseudoobscura* (Richards, Liu et al. 2005), *D. miranda* (Zhou and Bachtrog 2012), and the *de-novo* assembled *D. persimilis*, *D. lowei* , and *D. affinis*) to date the origin of the orphan genes to different ancestral nodes in the phylogenetic tree of these species (Beckenbach, Wei et al. 1993). We distinguished five groups of genes: old genes (non orphans) and four different orphan age classes (Figure 6). Surprisingly, we observed a consistent paucity of orphans on XR relative to XL across all age classes (Figure 7). This persistent difference in orphan number between XL and XR in all age classes suggests that X-linkage is not sufficient to explain the enrichment of orphans on XL. We conclude that the former autosome differs from the ancestral X chromosomal arm by a yet unidentified feature that affects the emergence of new orphans.

The analysis of orphans that have putatively lost their function *via* the acquisition of a stop codon or a frame shift causing insertion/deletion (pseudogenized/lost orphans) reveals another interesting feature of the XL-XR fusion. The oldest orphans in our data set (age class 4) show a pronounced excess of pseudogenized orphans on XR in *D. affinis* and *D. miranda* (Figure 8A). This trend was not observed for orphans that emerged on XR after the XL-XR fusion (Figure 8B-C), nor for old genes (Figure 8D) and is not due to an increased rate of orphan gain on XR (Figure 9). Since the oldest orphans (age class 4) on XR are a mixture of autosomal (*i.e.*, before the fusion) and sex-chromosomal (*i.e.* after the fusion) orphans, we speculate that the high rate of pseudogenization of orphans on the XR may reflect the new X-linkage of previously autosomal orphans. A previous study (Meisel, Han et al. 2009) found that the XR chromosome has experienced a burst of gene duplications to autosomes after its creation. It is plausible that after the conversion of the XR from autosome to sex-chromosome, orphans might have been duplicated to autosomes, while the XR ancestral copy would have become pseudogenized. To test this hypothesis, we looked for evidence of gene duplications for the orphans lost on the XR at node 4 (Figure 6). We aligned the sequences of these genes in *D. lowei* and *D. miranda* to the respective genomes using BLASTN (cutoff $10^{-5}$). Upon manual inspection of the alignments we found that only 1 out of 21 genes in *D. miranda* (gene ID: GA23486) and 1 out o f 14 genes in *D. lowei* (gene ID: GA23807) had a second hit on an autosome covering at least 50% of the length of the query gene. Other genes either produced a single best hit on the XR chromosome or spurious short hits on other chromosomes (data not shown). Thus, we conclude that duplication of orphans cannot explain the excess of pseudogenized orphans on XR. Nevertheless, our analysis clearly indicates that the emergence of the neo-X chromosome influenced the orphan dynamics on XR, affecting rates of both gain and loss, thus we excluded this chromosome arm for our analyses of the rate of orphan turnover.

For each age class, we determined the number of pseudogenized orphans (Tautz and Domazet-Loso 2011). In the *D. persimilis* lineage, orphan pseudogenization can be studied for three different age classes. If orphans of all age classes were functionally equivalent, no difference in the rate of orphan pseudogenization would be expected. We observe, however, that the fraction of orphan pseudogenes decreases with age (Figure 10). The *D. miranda* lineage also shows a higher loss of young orphan genes. The relatively small number of observations, however, precludes statistical testing of this trend. Overall, orphan genes are lost significantly more often than old genes (Fisher's exact test, P = 3.3 × $10^{-8}$), consistent with the rapid turnover hypothesis. The unequal conservation of orphans of different age classes is also apparent after normalizing by coding sequence length (Figure 11), to account for the fact that longer coding sequences (CDS) have a greater chance of acquiring ORF-disrupting mutations. When looking at the distribution of premature termination codons (PTC) along the open reading frame (ORF) of all genes, we observed that PTCs are enriched at the beginning and at the end of the ORF (Figure 12), consistent with previous results in *D. melanogaster* (Lee and Reinhardt 2012) and *D. pseudoobscura* (Hoehn, McGaugh et al. 2012). Since ORF-disrupting mutations occuring at the end of the ORF might have little impact on gene function, we redefined pseudogenes by considering only ORF-disrupting mutations localized in the first half of the ORF and confirmed that orphans of age class 3 are lost more often than those of age class 4 (Figure 13). Age class 2 was intermediate, most likely not reflecting a biological phenomenon, but due to a high sampling variance associated with the small number of observations (9 orphans). Finally, the pattern is also robust to a more conservative criterion for ortholog assignment (see Methods, section "Annotation of the *obscura* species" , Figure 14).

To determine features associated with the differences in disabling mutations among orphans from different age classes, we contrasted orphans lost in *D. lowei* and/or *D. persimilis* (lost orphans) *vs.* orphans conserved in all the *obscura* species (conserved orphans). Genes in both classes evolve at the same rate, are of similar length, and have similar codon usage bias (Figure 15A-C-E). Conserved orphans have a higher GC content, contain fewer microsatellites, are expressed at a higher level and are more male-biased (Figure 15B-D-F-G-H) compared to lost orphans. Conserved orphans tend to increase their expression level as they become older (Figure 16A), while the opposite pattern is true for lost orphans (Figure 16B).

Orphan genes are frequently expressed in the testis (Levine, Jones et al. 2006) (Begun, Lindfors et al. 2007) and have a male-biased gene expression pattern (Metta and Schlötterer 2008). This pattern could be generated by pervasive gene expression in testis, which facilitates the functional recruitment of non-specific expression (Kaessmann 2010). Another explanation is that expression in testis does not require a complex architecture of regulatory modules (Sassone-Corsi 2002; Kleene 2005; Kaessmann 2010), so that fewer substitutions are required to obtain a functional regulatory module for expressing a novel gene in testis compared to other tissues. We scrutinized these explanations by comparing the fraction of male-biased genes among orphan genes from different age classes. Unexpectedly, the fraction of male-biased genes increases with age of the orphan genes (Figure 17). This increase of male-biased orphans among the older age classes is the result of a preferential loss of orphans with an unbiased gene expression (Figure 18). To confirm that male-biased gene expression is associated with orphan retention rather than emergence, we analyzed the sex-bias in *D. miranda* for orphans with and without an open reading frame. Consistent with the gene expression pattern in *D. pseudoobscura*, we found that lost orphans have a significantly lower male-bias in *D. miranda* (Figure 19). We conclude that the previously reported male-biased gene expression of orphan genes is not the result of a preferential recruitment of male-biased transcripts, nor do orphans gradually acquire male-biased gene expression. Rather, male-biased orphans are more likely to be retained.

## Discussion

Our study provides the missing link to understand orphan dynamics. Until now, orphan evolution was primarily studied on long phylogenetic branches. While this approach is well suited to discover new orphans, it does not allow tracing the evolution of orphans. Previous studies showed a high rate of orphan gain, which is not reflected in an increase in gene number. To resolve this apparent paradox, it has been postulated that orphans must be lost at a high rate as well (Tautz and Domazet-Loso 2011). Here, we used the framework of closely related species in the *obscura* group to study the patterns of orphan gain and losses. We show that orphans do not only emerge at high rates, but that they are also rapidly lost (Figure 10). Interestingly, most losses (~76%) were due to disabling mutations rather than deletions of the orphan gene. While under equilibrium conditions the number of losses balances the number of orphan gains, here, we observed a surplus of orphan gains (Figure 6). We caution that this discrepancy probably does not imply an increase of gene number, but rather reflects the limited evolutionary time to acquire mutations. Using a rather conservative criterion for disabling mutations, either premature stop codons or frameshift indels, we have probably not identified all orphans that have lost their function. Furthermore, we do not account for the possibility of loss of function due to changes in gene regulation.

Importantly, codon usage bias, *dN/dS* values and sequence conservation clearly suggest that orphan genes are functionally constrained and this constraints do not differ among orphans that are conserved in the *obscura* group and those that lost function in at least one species of the group. Hence, it may be possible that orphan loss is stochastic and reflects weak purifying selection. Nevertheless, lost orphans differ in some aspects from conserved ones. Orphans that are lost contain more microsatellite stretches and have a lower, less sex-biased gene expression than retained ones. Furthermore, we also found that the rate of orphan loss decreases with orphan age, a result consistent with orphans serving a functional role only temporarily. Previous work suggested that orphans are important for adaptation to novel environments (Khalturin, Hemmrich et al. 2009; Colbourne, Pfrender et al. 2011), but it is also possible that orphans contribute to stabilize new connections in gene networks (Capra, Pollard et al. 2010; Warnefors and Eyre-Walker 2011) and become obsolete once such new connections have been optimized. Our data suggest that orphans become quickly functional, which is reflected in their codon usage bias, *dN/dS* ratio and sequence conservation.

The chromosomal translocation resulting in the neo-X chromosome provides another interesting perspective on the evolution of orphan genes. Despite the fact that the neo-X is now fully dosage compensated (Abraham and Lucchesi 1974), and has obtained a similar base composition as the XL (Gallach, Arnau et al. 2007), we noted that the translocation resulted in a preferential loss of orphan genes on the neo-X. Since this pattern is restricted to orphans that most likely originated before the chromosomal fusion, we argue that the change in chromosomal environment has affected the function of orphan genes, most likely via expression differences. We speculate that the selective advantage conferred by these orphans has diminished, which resulted in a higher loss rate. Interestingly, the elevated rate of orphan loss after the neo-X formation seems to be still ongoing. This differential loss of orphan genes point in a similar direction as the observation that the gene composition of the neo-X has been altered by gene duplication (Meisel, Han et al. 2009). Hence, both (orphan) gene loss and duplication contribute to fast gene content remodelling on a newly formed sex chromosome in *Drosophila*.

## Materials and methods

**Species data collection**

An individual species sample of *D. affinis* (stock number 14012‑0141.02) was ordered from the Drosophila Species Stock Center (https://stockcenter.ucsd.edu/info/welcome.php) and sequenced on the Illumina GAIIx following the paired-end library preparation protocol (version Illumina 1.7) in two runs (run 1: read length = 101 bp, insert size = 230 bp; run 2: read length = 101 bp, insert size = 550 bp). Short genomic reads *D. lowei* (accessions SRX091466 and SRX091467) and *D. persimilis* (accession SRX091471) were downloaded from the Short Reads Archive (http://www.ncbi.nlm.nih.gov/sra). The genome of *D. miranda* was downloaded from NCBI (GenBank Assembly ID GCA_000269505.1). The genome of *D. pseudoobscura* was downloaded from FlyBase (release 2.23).

**Assembly of the *obscura* species**

Reads for *D. affinis*, *D. lowei* and *D. persimilis* were trimmed using the Perl script trim_fastq.pl (parameters –quality-threshold 20 ––min-length 40) from PoPoolation (Kofler, Orozco-terWengel et al. 2011). For each species, a *de novo* assembly (parameters: min-contig-length 200) was performed using CLC Genomics Workbench 4.6 (http://www.clcbio.com/products/clc-genomics-workbench/), followed by scaffolding with nucmer (parameters: –c 30 –g 1000 –b 1000 –l 15) against the *D. pseudoobscura* genome. Average coverage per assembled genome was calculated by realigning the reads against the contigs of the respective species with Bowtie 2.1.0 (parameters: --very-fast) and selecting only reads with mapping quality > 20.

**Annotation of the *obscura* species**

The annotation of *D. affinis*, *D. lowei*, *D. miranda* and *D. persimilis* is based on orthology to *D. pseudoobscura* using Exonerate 2.2.0 (parameters: -model protein2genome –bestn 1 -showtargetgff), by aligning the longest isoform of *D. pseudoobscura* proteins extracted from a recent re-annotation of *D. pseudoobscura* (Palmieri, Nolte et al. 2012) to the genomes of *D. affinis D. lowei*, *D. miranda* and *D. persimilis*. For each gene, the best unambiguous hit was retained. To remove non-informative hits, we also required a minimum fraction of the gene to be recovered. Since the sequence conservation of orthologs decreases with divergence time, the expected length of the ortholog depends strongly on the phylogenetic distance between query and subject sequence. To apply consistent criteria for all species, we empirically determined the expected fraction of a gene with sequence homology. Based on genes that are conserved between *D. pseudoobscura* and the 10

*Drosophila* species outside the *obscura* clade (old genes) (Supplementary File 1) we determined the distribution of the fraction of the genes that could be aligned. As cutoff the value we used the 5$^{th}$ percentile of the distribution of aligned protein length of old genes. This resulted in a threshold of 47% for *D. affinis*, 52% for *D. lowei*, 59% for *D. miranda* and 53% for *D. persimilis*. Hence, only orphan orthologs that showed a fraction of aligned coding sequence higher than the empirically determined cutoffs were retained. In addition to this ortholog set, we generated an alternative, more conservative ortholog set. For this one, at least one of the flanking genes of *D. pseudoobscura* was required to be in synteny with the respective orthologs in *D. affinis*, *D. miranda* and *D. persimilis*. *D. lowei* was not considered in the synteny analysis since most of the genes in this species are flanked by genomic gaps, due to the shorter contig length of the *D. lowei* assembly (Supplementary File 2) which caused many contigs to contain only a single gene (Supplementary File 3), thus precluding proper synteny assignments. Assembly and annotation of all the species are available at http://popoolation.at/affinis_genome, http://popoolation.at/lowei_genome, http://popoolation.at/miranda_genome and http://popoolation.at/persimilis_genome. Detailed annotation statistics for each gene are reported in Supplementary File 4.

**Detection of orphan genes**

*D. pseudoobscura* proteins corresponding to the longest isoform for each gene were aligned using BLASTP (E< 10$^{-4}$) and TBLASTN (E< 10$^{-4}$) against the published proteomes and genomes of 10 *Drosophila* species outside the *obscura* group (*D. melanogaster, D. simulans, D. sechellia, D. erecta, D. yakuba, D. ananassae, D.*

*willistoni*, *D. mojavensis*, *D. virilis D. grimshawi*). Genes without BLAST hits and without annotated orthologs in FlyBase (gene orthologs release 09-2011) were classified as orphans.

**Polymorphism analysis**

Illumina reads for 45 *D. pseudoobscura* strains were downloaded from NCBI (Short Reads Archive, accession SRP017196). Reads were trimmed using PoPoolation (Kofler, Orozco-terWengel et al. 2011) and a total of 3.5 million reads was randomly extracted for each strain and combined into a single FASTQ file. The combined reads were treated as a Pool-Seq data set and mapped to the FlyBase *D. pseudoobscura* genome release 2.23 with BWA (Li and Durbin 2009) (parameters -o 1 -n 0.01 -l 200 -e 12 -d 12) on a hadoop cluster using DistMap (Pandey and Schlötterer 2013). From the resulting BAM file, PCR duplicates were removed with Picard (http://picard.sourceforge.net) using the tool MarkDuplicates.jar (parameters REMOVE_DUPLICATES=true, VALIDATION_STRINGENCY=SILENT). Proper-pairs with mapping quality > 20 were extracted with samtools (version 0.1.18) (Li, Handsaker et al. 2009). Indels were detected with PoPoolation using the script identify-genomic-indel-regions.pl (parameters --min-count 2 --indel-window 5) and masked from the reference genome prior to SNP calling. Coverage was subsampled to 50X for all the chromosomes. Only SNPs on the 3$^{rd}$ chromosome were considered in all analyses, since a balancer chromosome was used to extract the 3$^{rd}$ chromosome, precluding an unbiased polymorphism analysis for the remaining chromosomes. SNPs were called with the PoPoolation script Variance-sliding.pl (parameters --min-coverage 10

--min-count 2 --max-coverage 500 --min-qual 20 --window-size 500 --step-size 500 --fastq-type sanger --pool-size 45).

**Calculation of orphan gains and losses**

Orphan gains and losses (pseudogenizations) were inferred by Dollo parsimony. Based on the phylogenetic tree of Figure 6, a gene was assigned as gained at a given node if an intact ortholog was present in both external branches of the subtree corresponding to that node. For example, a gene having an intact ORF in *D. lowei* but not in *D. affinis* was classified as gained at node 3 (Figure 6). A gene was considered to be lost at a terminal branch if at least one ORF-disrupting mutation (frameshift / premature stop codon) was present in the gene at that branch and two intact ORFs were detected at both external leaves (Wang, Grus et al. 2006). The relatively high coverage of our assemblies (Supplementary File 2) makes unlikely that disrupting mutations are sequencing errors. In *D. affinis* for instance, only 8 genes had an average coverage lower than 20x.

A gene was considered as completely deleted in a species if no ortholog was detected in that species and no BLASTP ($E< 10^{-4}$) or TBLASTN ($E< 10^{-4}$) hit was found. Deletions were not considered into analyses of gene turnover, since they cannot be distinguished from missing annotations.

**Expression analysis**

Four RNA-Seq data sets of *D. pseudoobscura* males and females (strains ps94 and ps88 from the ArrayExpress database – accession E-MTAB-1424) together with two RNA-Seq samples of *D. miranda* males and females from the Short Reads Archive (accessions SRX106024, SRX106025) were used for expression analysis. For each sample, reads were trimmed using PoPoolation (Kofler, Orozco-terWengel et al. 2011) and aligned to the genome of the respective species with GSNAP version 2012-07-12 (Wu and Nacu 2010) (parameters: –N 1). Only proper pairs mapping unambiguously to one position were retained. Expression in FPKM was calculated with Cufflinks version 1.2.1 (parameters: -F 0.10 –j 0.15 –I 300000). For *D. pseudoobscura* sex-bias was calculated using the package DESeq (Anders and Huber 2010), treating the strains as two biological replicates for each sex and applying an FDR = 0.1. Differential expression between *D. miranda* males and females was calculated for both species using the $\log_2$ fold change on the normalized expression counts using the normalization protocol implemented in the R package DESeq (Anders and Huber 2010) version 1.10.1.

**Codon usage bias**

Codon usage bias was calculated using the R package seqinr (function cai) based on the *D. pseudoobscura* codon usage table downloaded from *http://www.kazusa.or.jp/codon/cgi-bin/showcodon.cgi?species=7237*.

**Evolutionary rates**

Coding sequences of *D. pseudoobscura* and *D. miranda* orthologs without frameshifts / stop codons were aligned using PRANK (Loytynoja and Goldman 2005) (parameters: –codon). To test for purifying selection on orphans, *dN/dS* was compared between orphans and a set of randomly selected intergenic regions. This set was generated as follows: 1) we identified the intergenic regions from the *D. pseudoobscura* annotation from (Palmieri, Nolte et al. 2012), 2) for each CDS belonging to an orphan gene we extracted all the intergenic regions longer than that CDS, 3) we randomly selected one intergenic region and a random subregion with the same length of a given orphan CDS 4) this procedure was repeated for all orphan CDS, resulting in a set of intergenic regions with the same length distribution as orphan CDSs. These regions were aligned with BLASTN (cutoff $10^{-5}$) to the *D. affinis* genome and for each region the best hit was kept and realigned with PRANK (default parameters) to the *D. pseudoobscura* query sequence. Each alignment was truncated at the 5'end to get an alignment length which is multiple of 3. Internal stop codons were replaced by Ns. The ratio of the rates of nonsynonymous and synonymous substitutions per gene (*dN/dS*) were measured using Markov models of codon evolution and maximum likelihood methods implemented in PAML (Yang 2007).

**Comparison of genomic features among old-X, neo-X and autosomes**

To shed light on the differences in orphan number between XL and XR, different features were compared among old-X, neo-X and autosomes in *D. pseudoobscura* (unassembled contigs were not considered in this analysis): A) GC content was calculated with the R package seqinr for 10 kb and 100 kb sliding windows along each chromosome B) microsatellite density was calculated using SciRoKo 3.4 (Kofler, Schlötterer et al. 2007) (parameters: -mode mmfp –l 15 –r 3 –s 15 –p 5 –seedl 8 –seedr 3 –mmao 3) for 10 kb and 100 kb sliding windows along each chromosome; C) transposon density was estimated with RepeatMasker 3.2.9 (parameters: –q –gff -nolow –norna –species drosophila) for 10 kb and 100 kb sliding windows along each chromosome; D) length of intergenic regions were calculated using BEDTools (-complement) by interval subtraction between genome and gene coordinates; E) recombination rates for different windows were taken from McGaugh et al. (McGaugh, Heil et al. 2012).

**Microsatellite detection**

Microsatellites were detected on the transcript sequences of the longest isoform for each *D. pseudoobscura* gene using the tool SciRoKo (Kofler, Schlötterer et al. 2007) 3.4 (parameters: -mode mmfp –l 15 –r 3 –s 15 –p 5 –seedl 8 –seedr 3 –mmao 3).

**Transposons detection**

Genomic annotation of transposons was performed in *D. pseudoobscura* using RepeatMasker 3.2.9 (parameters: –q –gff -nolow –norna –species drosophila). Only transposons longer than 50 bp and not overlapping with microsatellites (see Methods, section "Microsatellite detection") were retained. We required for an orphan to contain a full transposon sequence in one of its exons in order to classify it as associated with a transposon.

## Competing interests

The authors declare that they have no competing interests.

## Authors' contributions

CS conceived the study. CS, CK, NP wrote the paper, NP performed the analyses.

## Funding

This work was supported by an Austrian Science Funds (FWF) grant P22834 to CS.

## Acknowledgements

We thank Ram Vinay Pandey for help with programming and Viola Nolte for support with genome annotations and library preparation. We are grateful to the members of the institute, in particular A. Betancourt and R. Kofler, for valuable discussion and comments on the manuscript.

# Figures

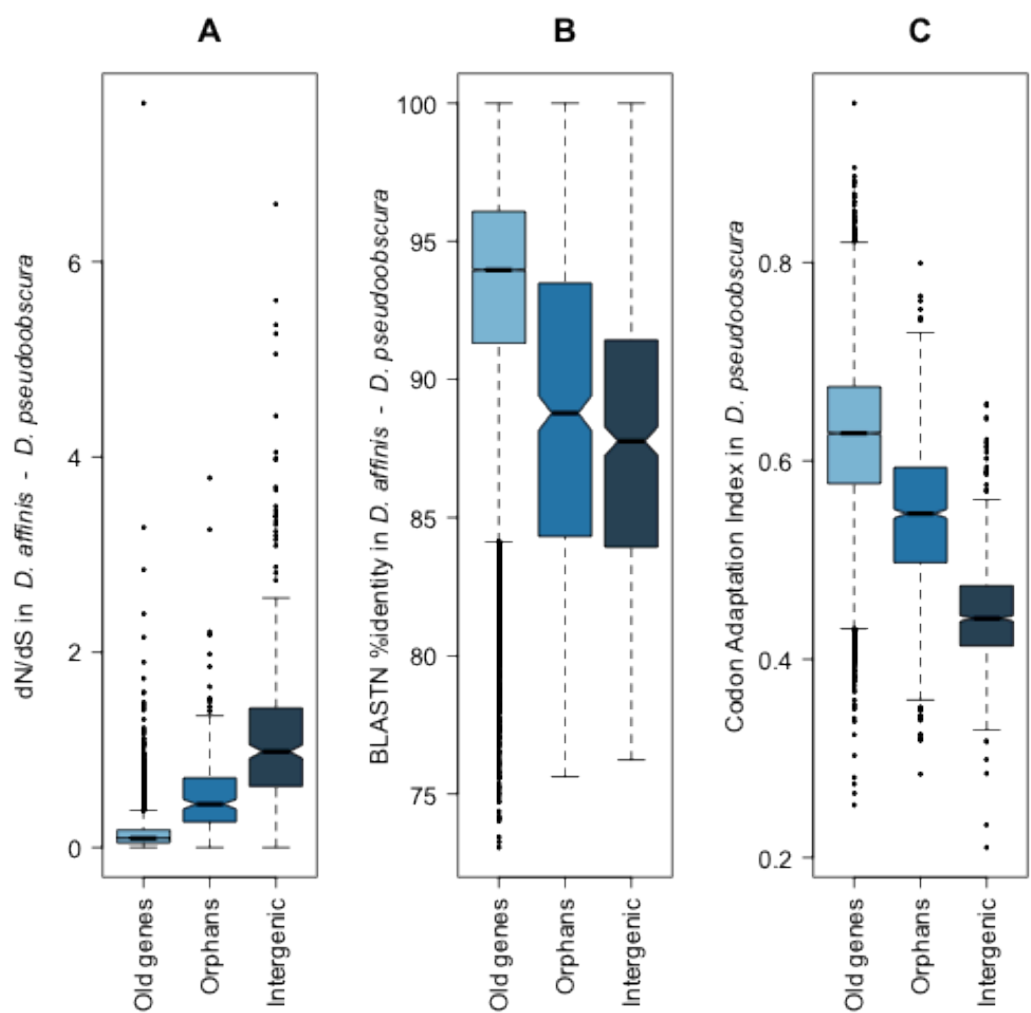

**Figure 1: Orphans are subject to purifying selection**

A) *dN/dS* of *D. pseudoobscura* and *D. affinis* orthologs. *dN/dS* is lowest for old genes, but also orphan genes have *dN/dS* smaller than one. A comparison of orphans and intergenic regions shows that *dN/dS* for orphans is significantly smaller (Mann-Whitney test, P = 9.5 × $10^{-10}$), indicating purifying selection on orphan genes. Intergenic regions were of similar length and chromosomal position as the orphan genes. B) Sequence similarity in HSPs obtained from BLASTing *D. pseudoobscura* genes against the *D. affinis* genome. Orphans are more conserved than intergenic regions (Mann-Whitney test, P = 0.00238) and less conserved than old genes (Mann-Whitney test, P < 1.0 × $10^{-15}$). C) Codon usage was measured by the Codon Adaptation Index (Sharp and Li 1987). The codon usage of orphans is significantly higher than that of intergenic regions (Mann-Whitney test, P < 1.0 × $10^{-15}$) indicating that orphans are subject to purifying selection. In comparison to old genes, orphans have a significantly lower codon usage bias (Mann-Whitney test – P< 1.0 × $10^{-15}$). Overall, all three analyses demonstrate that orphans are not annotation artifacts, but evolutionary conserved genes.

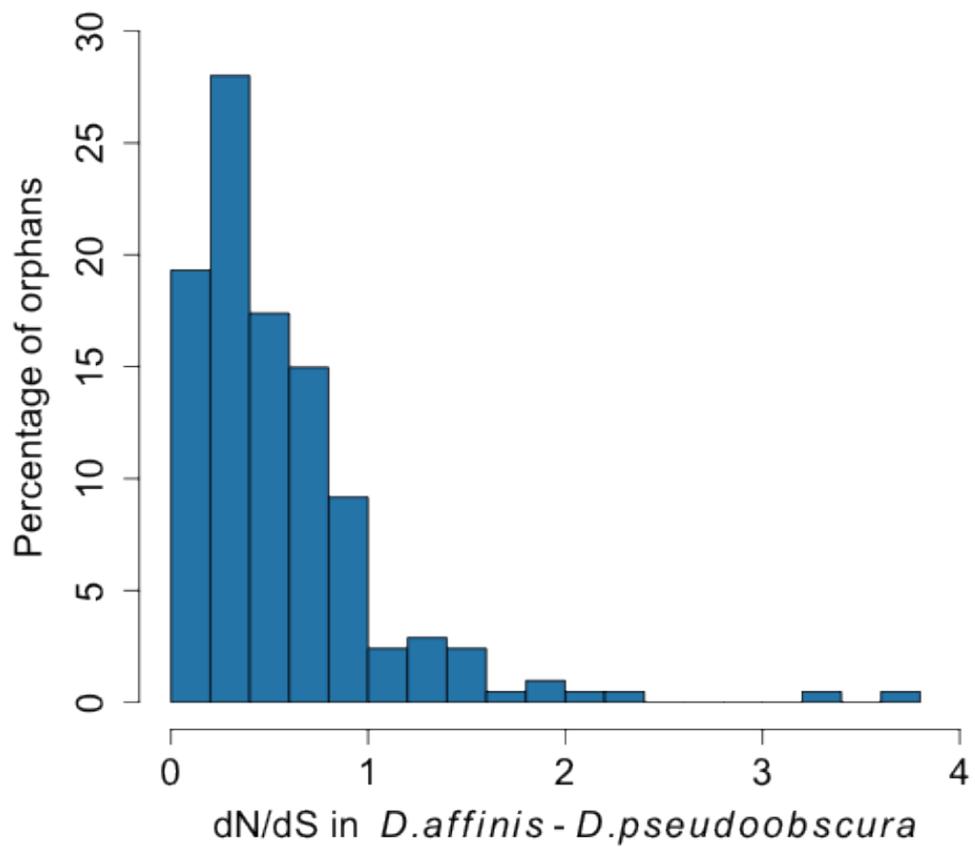

**Figure 1 – figure supplement 1: Distribution of *dN/dS* for orphan genes**
Most orphans have dN/dS lower than 1, consistent with the hypothesis of purifying selections acting on these genes.

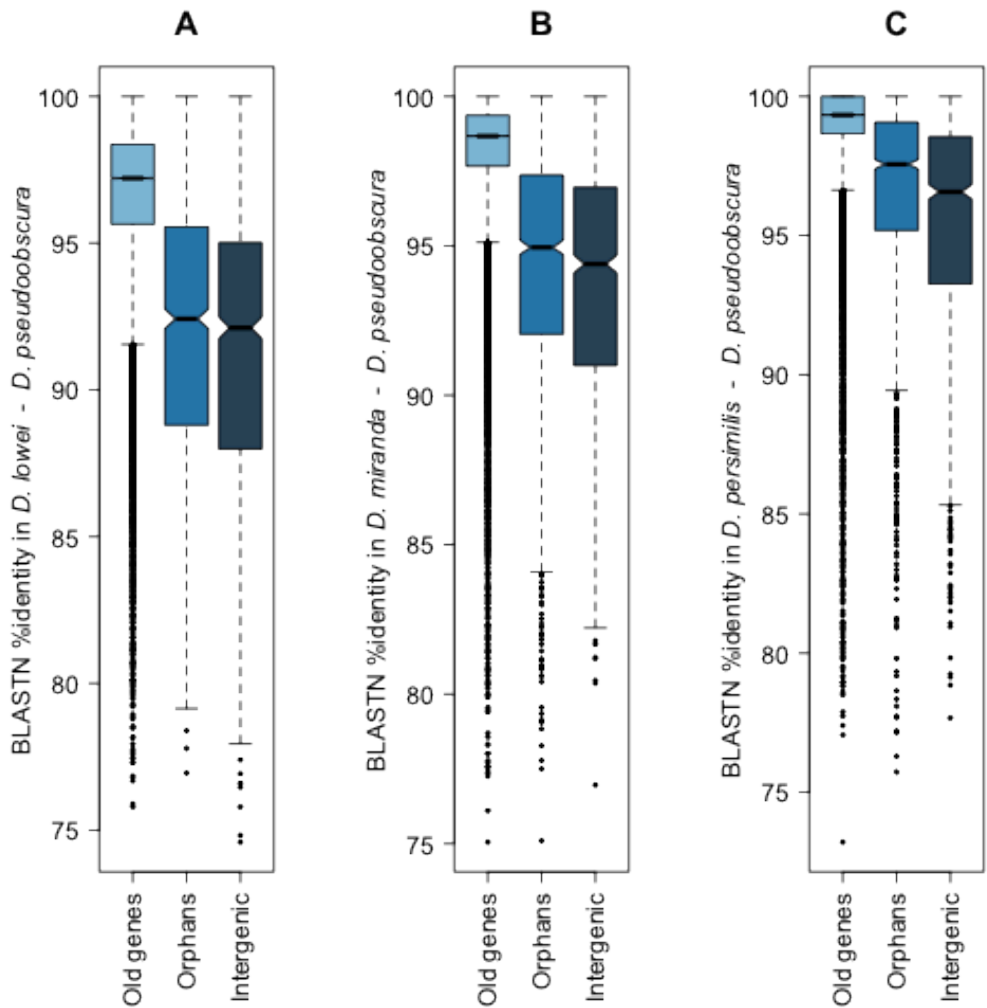

**Figure 1 – figure supplement 2: conservation of orphans in the *obscura* group**

Sequence similarity of old genes, orphans and random intergenic region obtained from BLASTing *D. pseudoobscura* genes against the genomes of *D. lowei* (A), *D. miranda* (B) and *D. persimilis* (C). Orphans are significantly more conserved than random intergenic regions in *D. lowei* (Mann-Whitney test, P = 0.00857), *D. miranda* (Mann-Whitney test, P < 0.00034) and *D. persimilis* (Mann-Whitney test, P < $2.8 \times 10^{-13}$). These results are consistent with purifying selection acting on orphans.

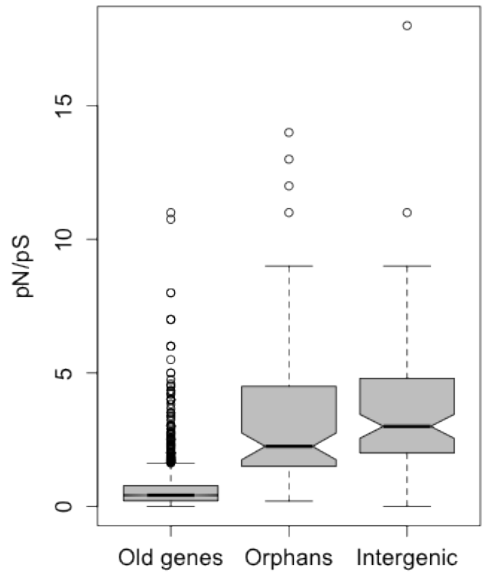

**Figure 2: *pN/pS* for old genes, orphans and intergenic regions**

Orphans show a *pN/pS* intermediate between old genes and intergenic regions. Nevertheless, *pN/pS* is significantly smaller for orphans compared to intergenic regions (Mann-Whitney test, P < 1.0 × 10$^{-15}$), indicating coding purifying selection acting on orphans.

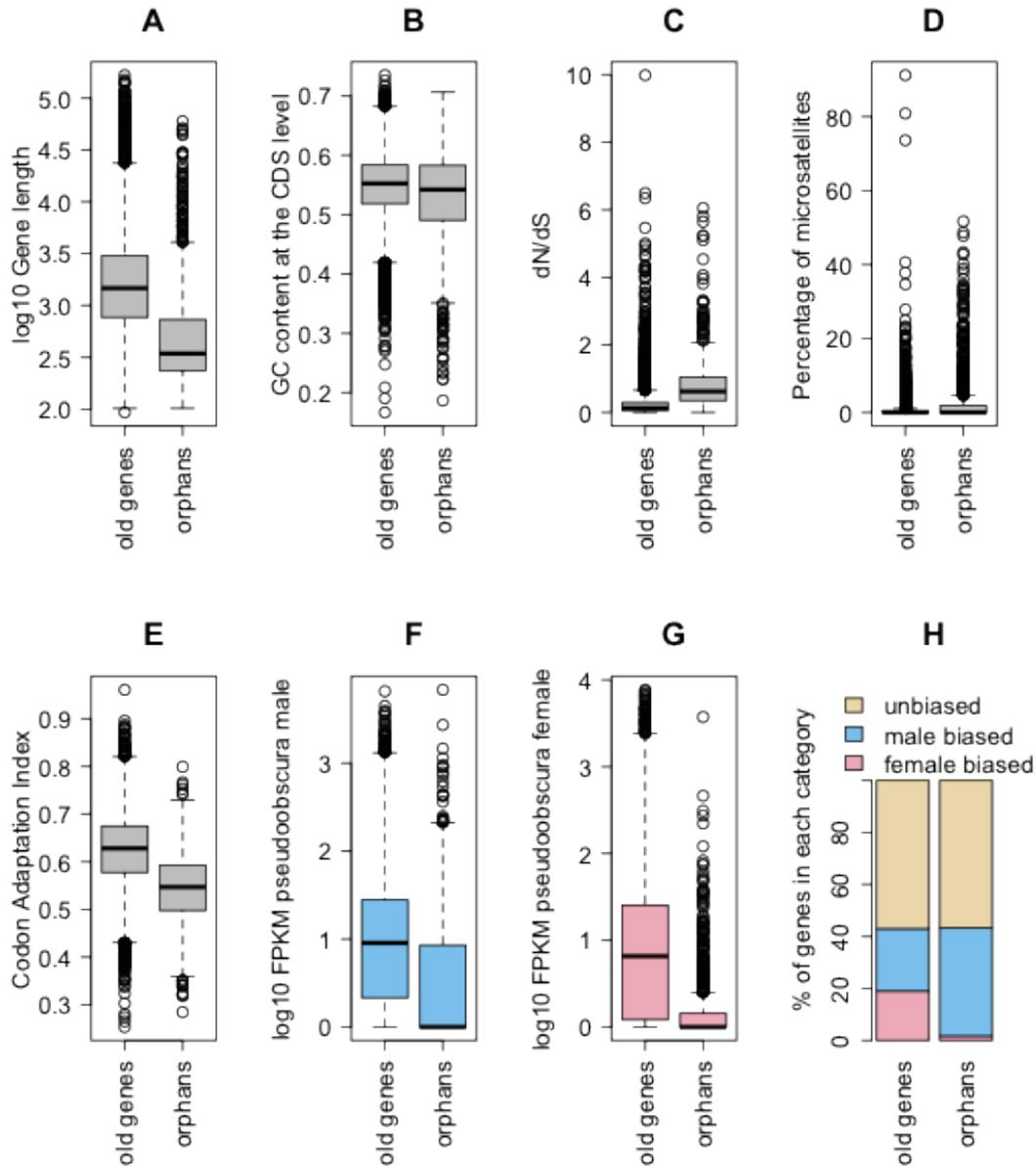

**Figure 3: Comparison of orphans and genes conserved among 10 *Drosophila* species outside of the *obscura* group**

Orphans differ from old genes in various features: A) gene length B) GC content, C) *dN/dS* D) percentage of microsatellites in coding sequence E) Codon Adaptation Index F) gene expression level in *D. pseudoobscura* males G) gene expression level in *D. pseudoobscura* females H) sex-biased expression. Orphans are shorter (Mann-Whitney test, $P < 1.0 \times 10^{-15}$), have lower GC content (Mann-Whitney test, $P = 3.9 \times 10^{-7}$), lower codon usage bias (Mann-Whitney test, $P < 1.0 \times 10^{-15}$), lower expression (Mann-Whitney test, $P < 1.0 \times 10^{-15}$), higher proportion of microsatellites (Mann-Whitney test, $P = 1.8 \times 10^{-4}$) and higher *dN/dS* (Mann-Whitney test, $P < 1.0 \times 10^{-15}$) compared to old genes. Moreover, orphans are more enriched in male-biased genes compared to old genes ($\chi^2$-test, $P < 1.0 \times 10^{-15}$).

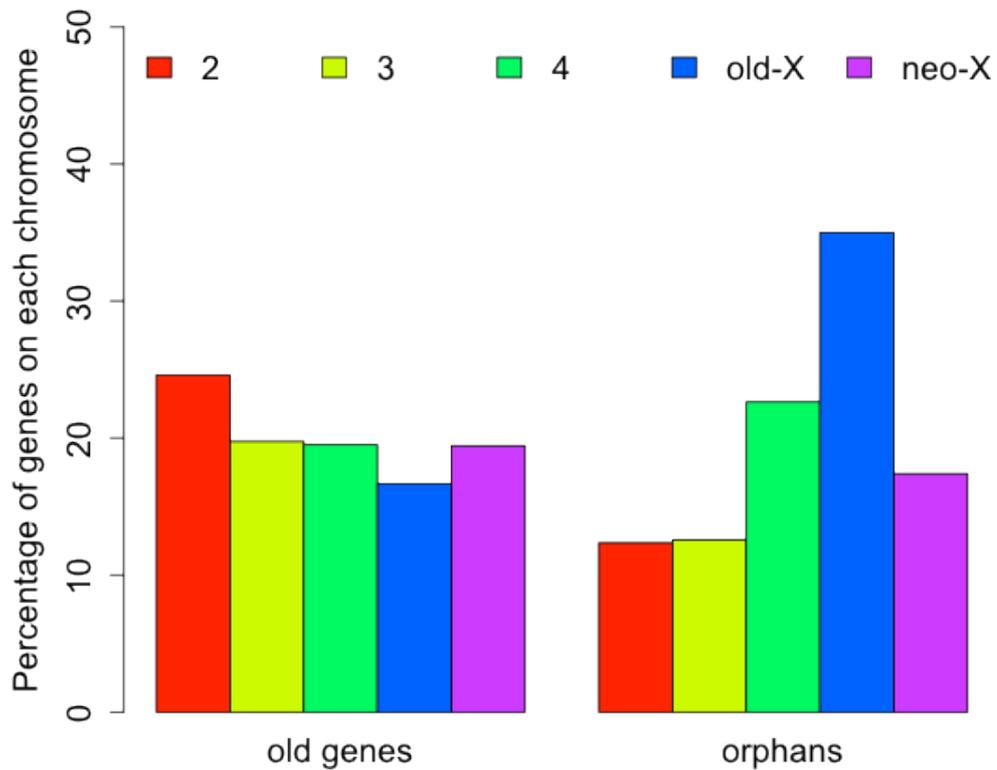

**Figure 4: Chromosomal distribution of old genes and orphan genes**

Orphans are overrepresented on the old-X. The number of orphan genes on the neo-X (XR) is significantly lower than on the old-X (XL) ($\chi^2$-test, $P < 1.0 \times 10^{-15}$).

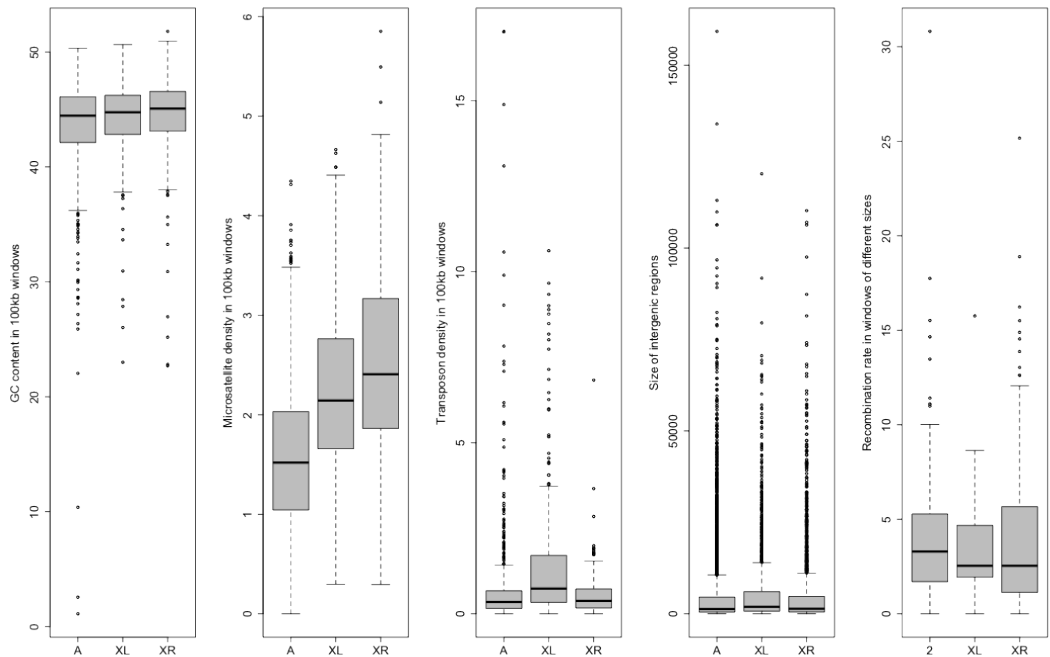

**Figure 5: Comparison of genomic features among autosomes, old-X and neo-X**

A) GC content in 100 kb windows, B) Microsatellite density in 100 kb windows, C) Transposon density in 100 kb windows, D) Length of intergenic regions, E) Recombination rate. GC content is significantly greater on the neo-X compared to old-X for 10 kb windows (Mann-Whitney test, P = 0.00020), but not for 100 kb windows (Mann-Whitney test, P = 0.1092). Microsatellite density is significantly higher on the neo-X for both windows of 10 kb (Mann-Whitney test, P = 1.9 × 10$^{-12}$) and 100 kb (Mann-Whitney test, P = 0.00025). Transposon density is significantly lower on the neo-X for both windows of 10 kb (Mann-Whitney test, P < 1.0 × 10$^{-15}$) and 100 kb (Mann-Whitney test, P = 4.6 × 10$^{-12}$). Intergenic regions are significantly shorter on the neo-X compared to the old-X (Mann-Whitney test, P = 7.4 × 10$^{-9}$). Recombination rate does not differ significantly between old-X and neo-X (Mann-Whitney test, P = 0.629).

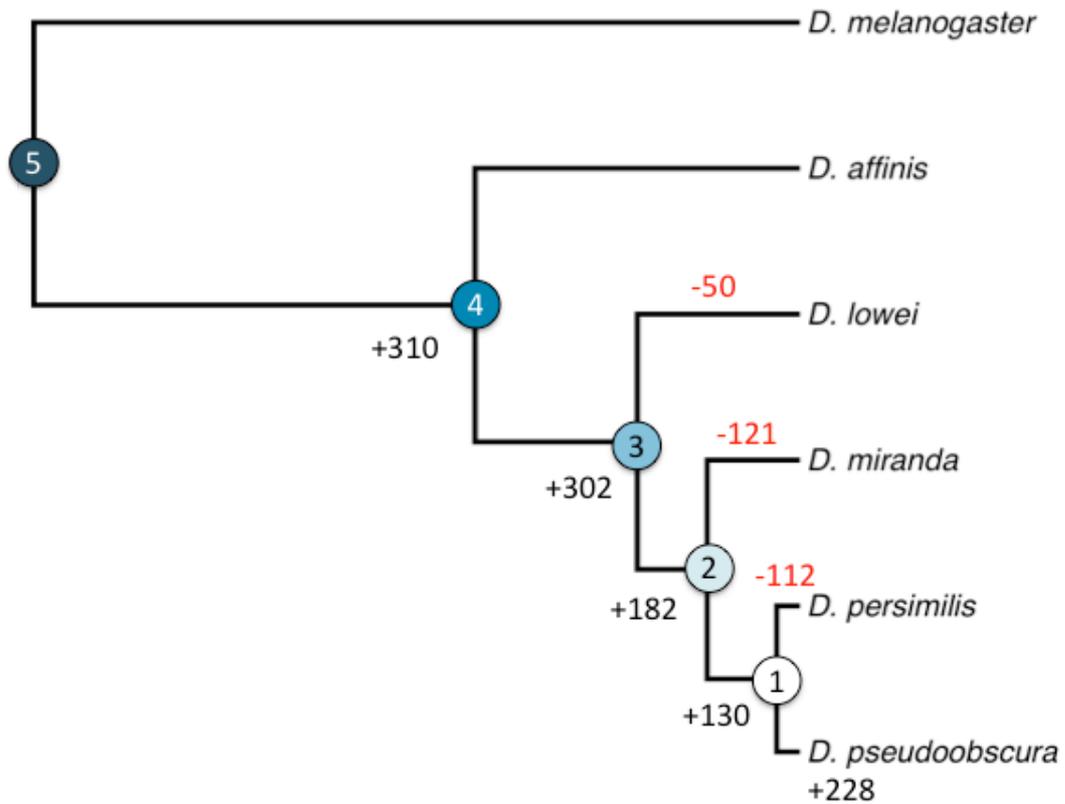

**Figure 6: Orphan gain and losses in the *Drosophila obscura* group**

A) Schematic phylogenetic tree of the *Drosophila obscura* group species according to Beckenbach et al. (Beckenbach, Wei et al. 1993) with *D. melanogaster* as outgroup. Genes conserved between *D. pseudoobscura* and 10 non-*obscura Drosophila* species correspond to age class 5 (old genes). For each age class the number of gene gains is shown in black. Orphans lost at a given branch are indicated in red. Note that losses at internal branches cannot be calculated, since all the orphans are present in *D. pseudoobscura*. Losses in *D. affinis* cannot be unambiguously assigned due to the absence of an additional *obscura* outgroup.

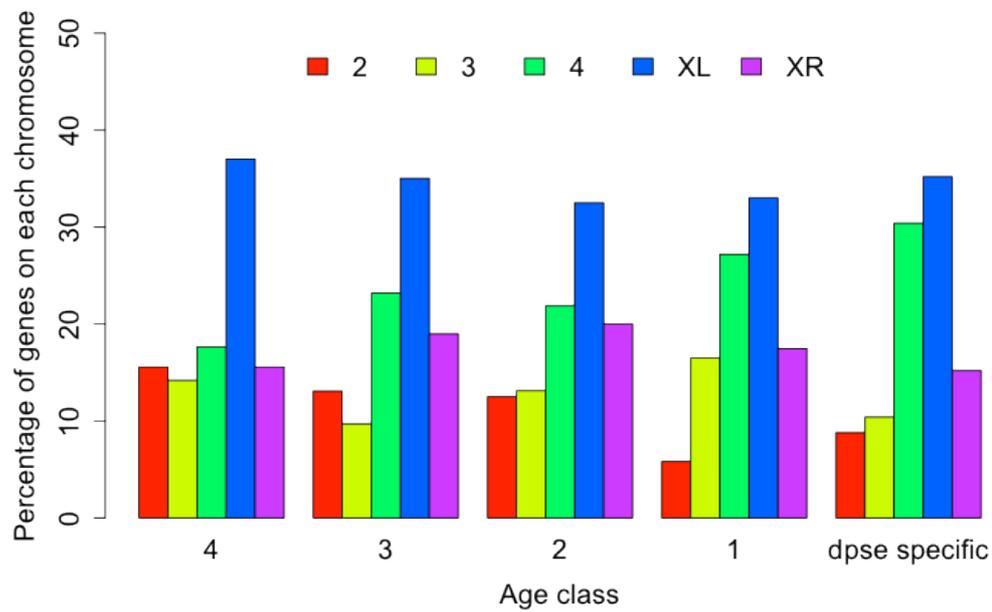

**Figure 7: Chromosomal distribution of orphans of different age classes**

In each age class orphans are underrepresented on the neo-X (XR) compared to old-X (XL) (Age class 4: $\chi^2$-test, P = 6.3 × $10^{-9}$; age class 3: $\chi^2$-test, P = 4.4 × $10^{-5}$; age class 2: $\chi^2$-test, P = 0.00590; age class 1: $\chi^2$-test, P = 0.00876; *D. pseudoobscura* specific: $\chi^2$-test, P = 0.00030).

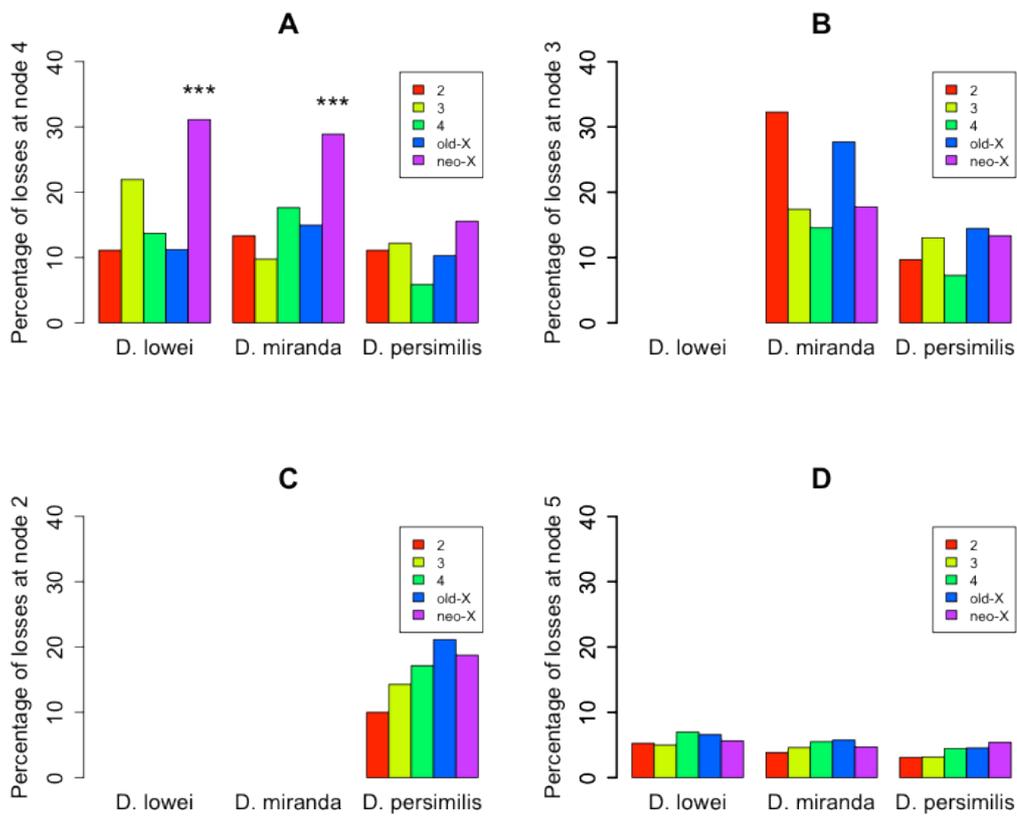

**Figure 8: Orphans predating the XL-XR fusion are preferentially lost on the neo-X**

For three terminal branches (*D. lowei*, *D. miranda*, and *D. persimilis*) the fraction of lost genes for each age class is shown. Each autosome and both X-chromosome arms are shown in different color. At node 4, where the neo-X originated, we observed the highest rate of orphan pseudogenization on the neo-X (A). Notably, this effect is not seen for younger orphans (B, C) neither for old genes (D).

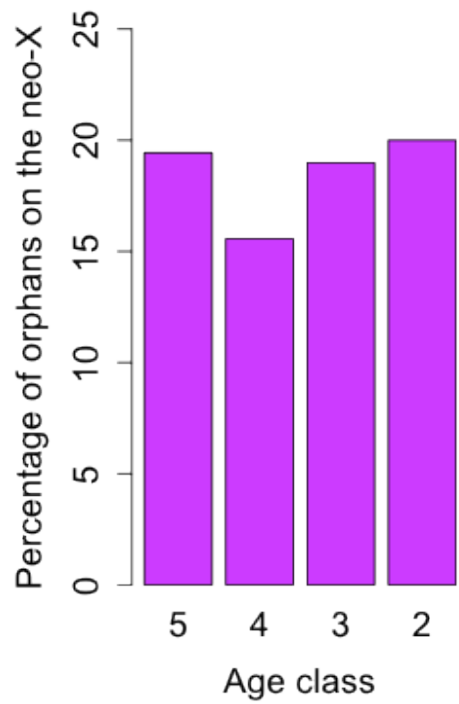

**Figure 9: No change in orphan gain on the neo-X chromosome**
The percentage of orphan genes on the neo-X chromosome remains constant through time (indicated by age classes).

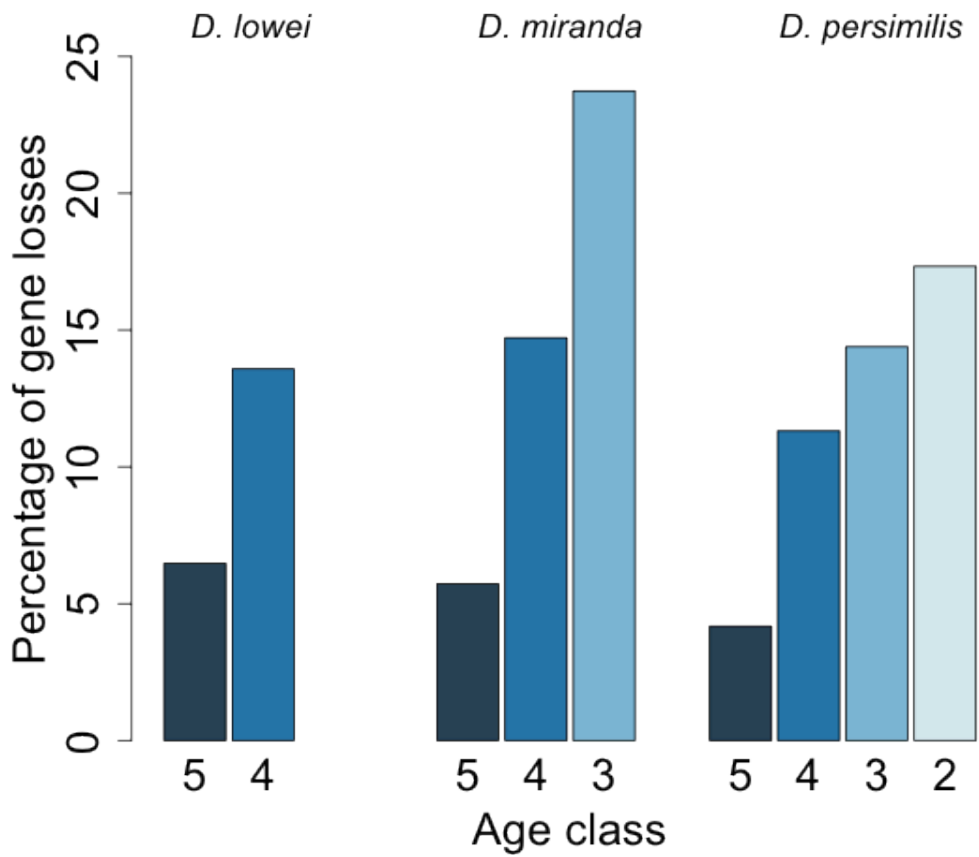

**Figure 10: Young orphan genes are more likely to be lost**

The barplot shows the fraction of orphans that has acquired a frameshift or premature stop codon (i.e.: lost function). For *D. lowei*, *D. miranda*, and *D. persimilis*, the fraction of lost orphans is shown for different age classes. Orphans are more likely to be lost than old genes. Both the *D. miranda* and *D. persimilis* lineage show that younger orphans are more likely to lose function than older ones.

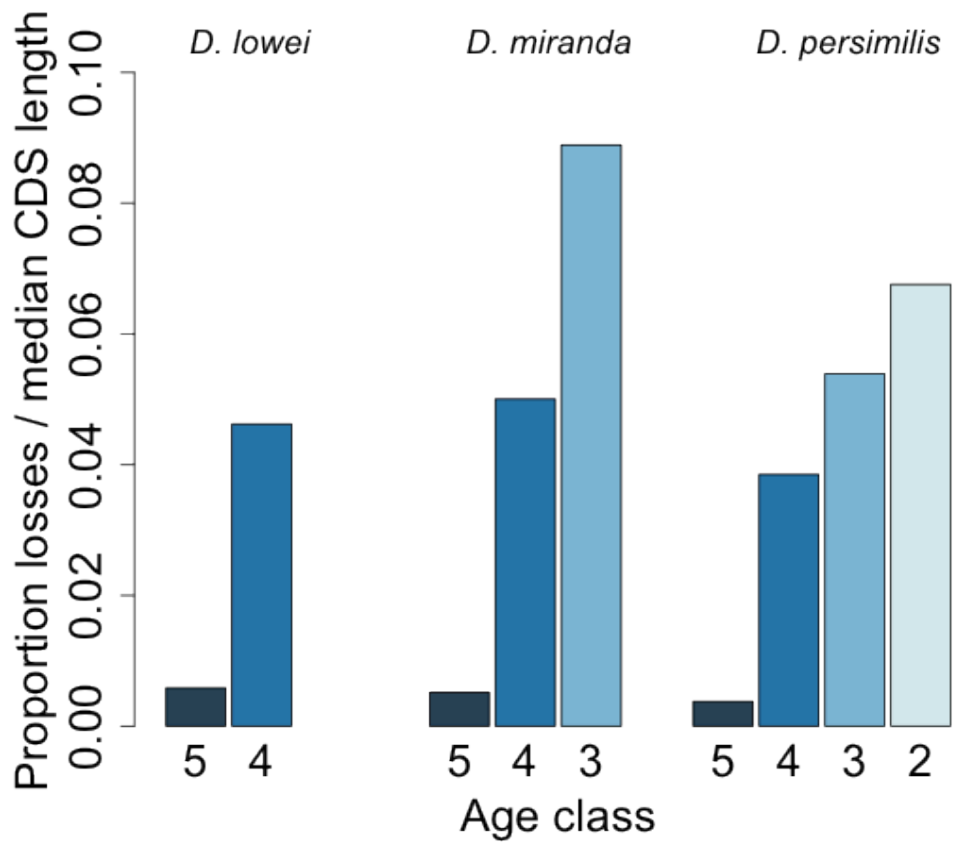

**Figure 11: Young orphan genes are more likely to be lost: accounting for CDS length**

To test if the short CDS of orphans affects the pattern that young orphans are more likely to lose function, we normalized the percentage of losses by the median CDS length of genes at that node.

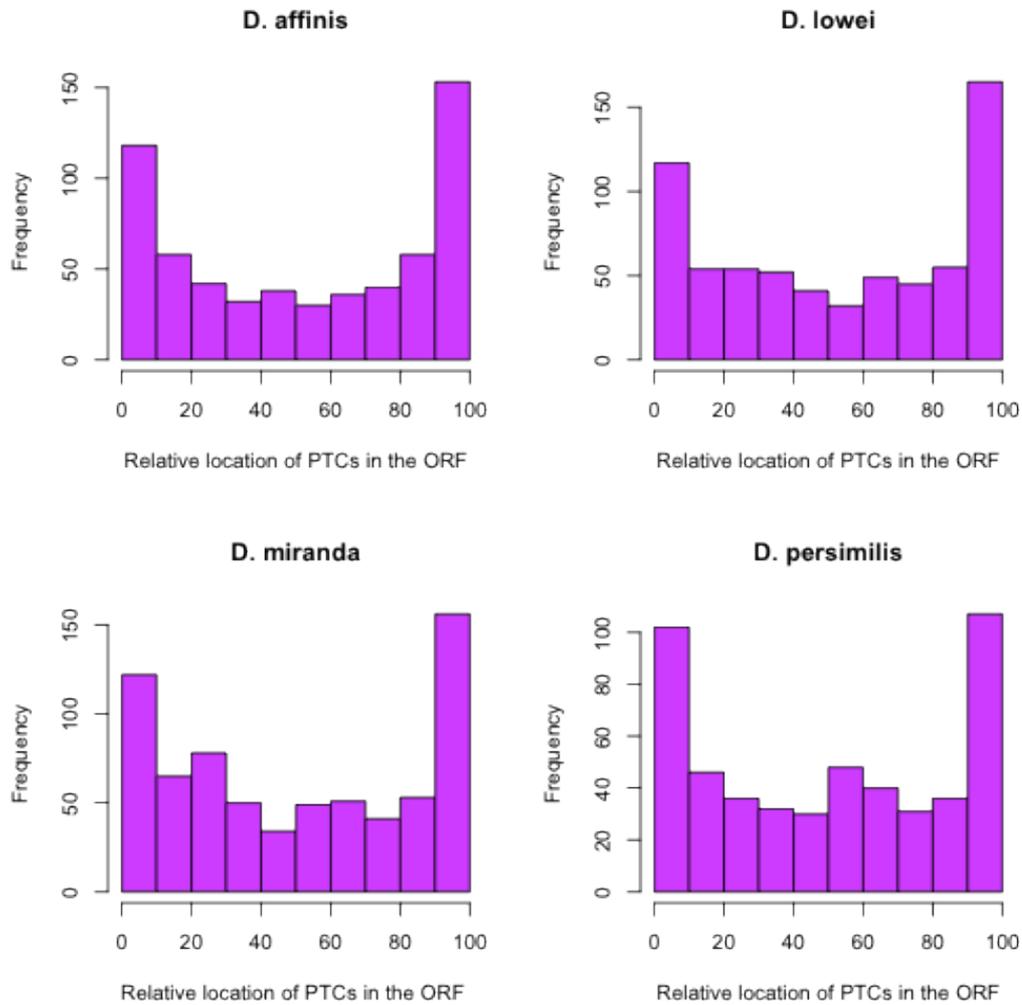

**Figure 12: Distribution of premature stop codons (PTCs) along the ORF for all genes containing PTCs**

PTCs are enriched at the beginning and at the end of the ORF in each species.

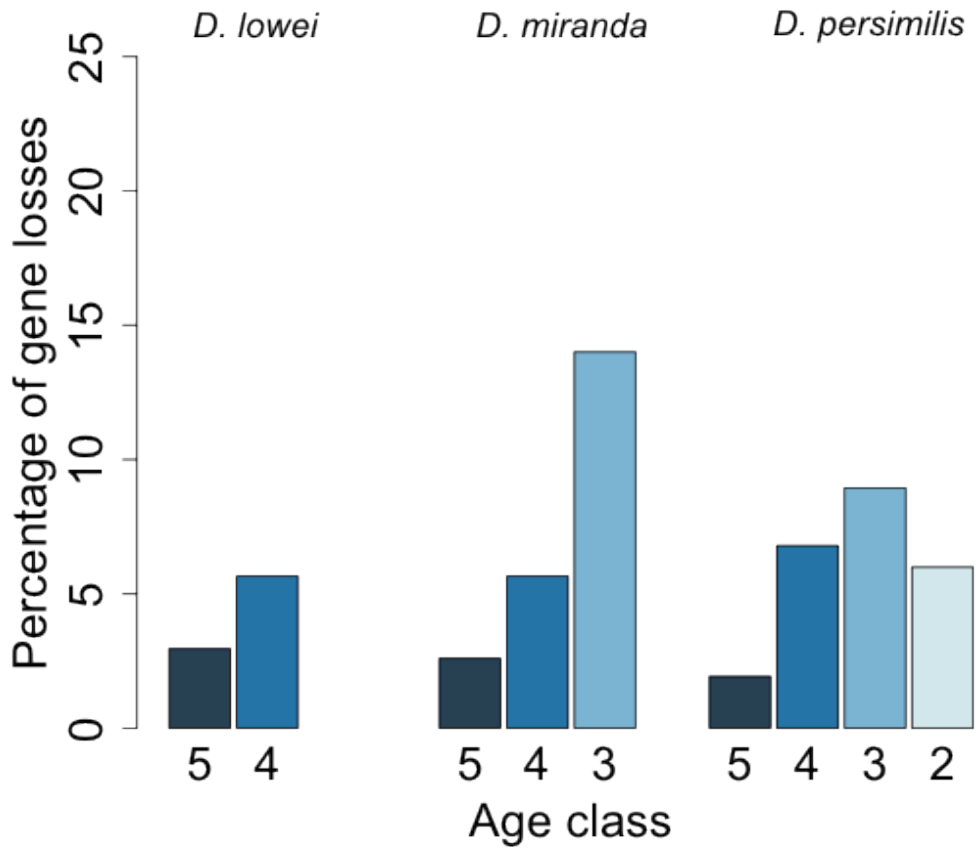

**Figure 13: Young orphan genes are more likely to be lost: considering only frameshifts and premature stop codons occuring in the first half of the ORF**

We repeated the analysis shown in Figure 10 by considering only frameshifts and premature stop codons occurring in the first half of the ORF to define a conservative set of pseudogenes, since disrupting mutations occurring at the end of the ORF are likely to have little impact on the gene function.

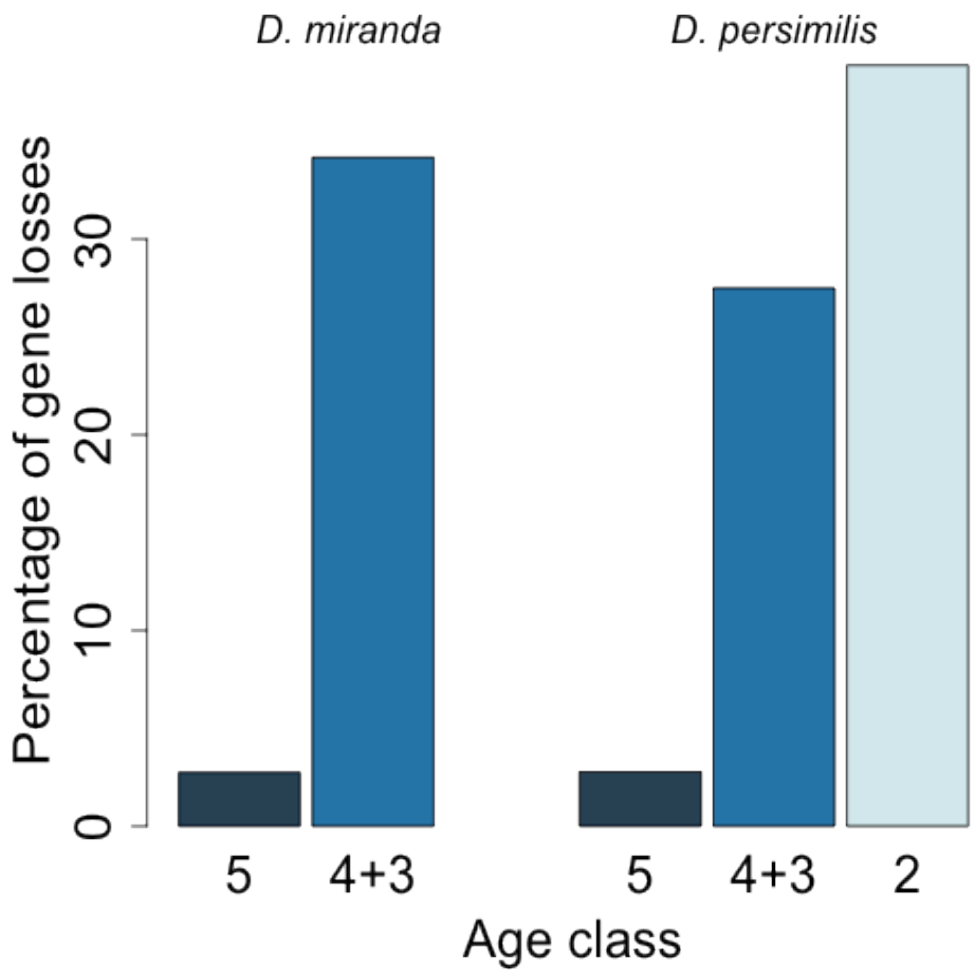

**Figure 14: Young orphan genes are more likely to be lost: the conservative set of orthologs**

We repeated the analysis shown in Figure 10 by restricting it to orthologs for which at least one flanking gene is identified in the same contig (see Methods, section "Annotation of the *obscura* species"). Due to the substantially reduced number of orphans in the older age classes, we combined age class 3 and 4.

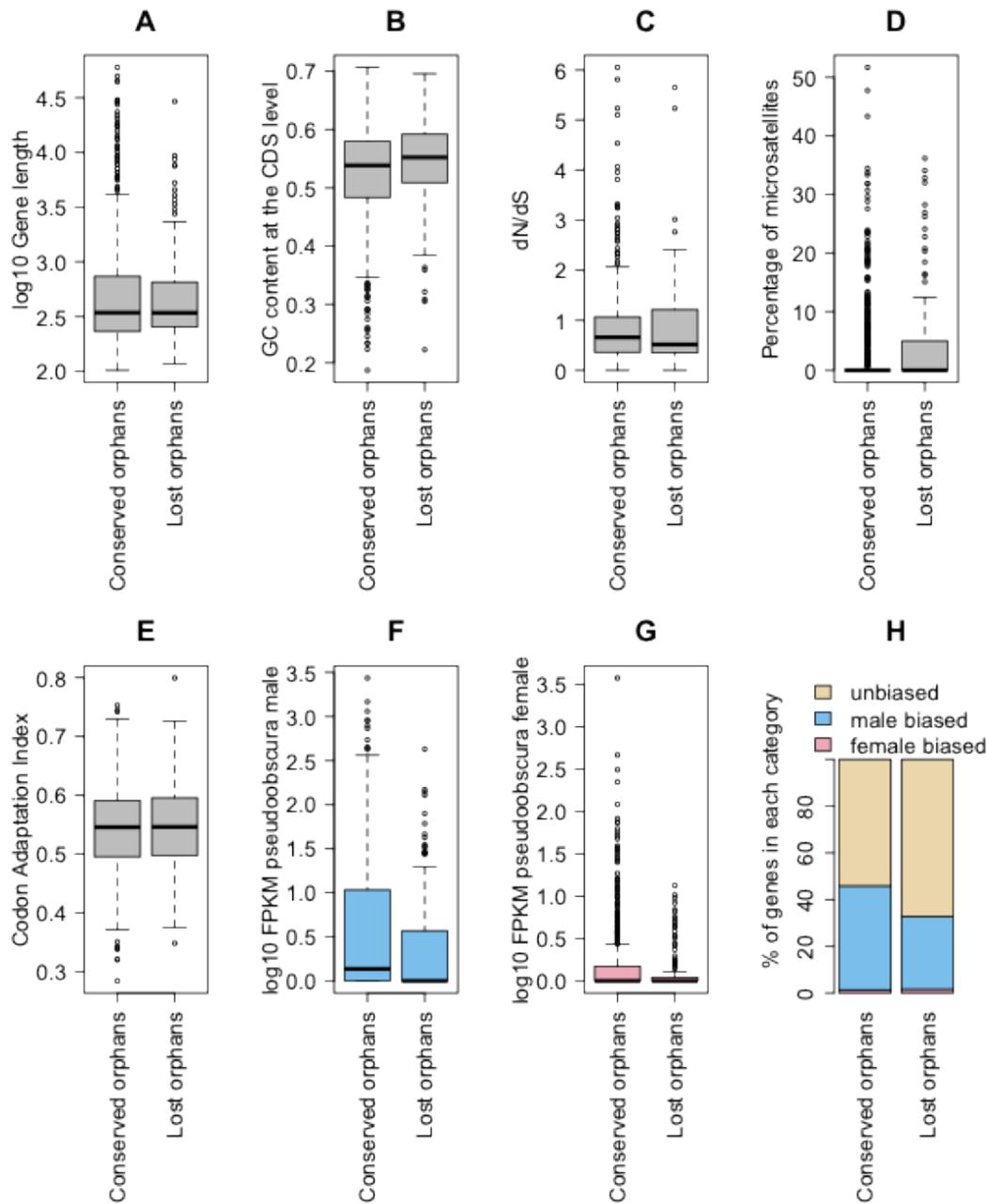

**Figure 15: Features of conserved orphans vs. lost orphans measured in *D. pseudoobscura***

A) Gene length B) GC content, C) *dN/dS* D) percentage of microsatellites in coding sequence E) Codon Adaptation Index F) gene expression levels in *D. pseudoobscura* males G) gene expression levels in *D. pseudoobscura* females H) sex-biased expression. Gene length (Mann-Whitney test, P = 0.7235) and evolutionary rates (Mann-Whitney test, P = 0.5835) are not significantly different between conserved and lost orphans. Lost orphans have higher GC content (Mann-Whitney test, P = 0.00325), lower expression in *D. pseudoobscura* males (Mann-Whitney test, P = 0.00012) and females (Mann-Whitney test, P = 0.00230) and a higher microsatellite content (Mann-Whitney test, P = 0.00049) compared to conserved orphans. Lost orphans are enriched in unbiased genes compared to conserved orphans ($\chi^2$-test, P = 0.02611).

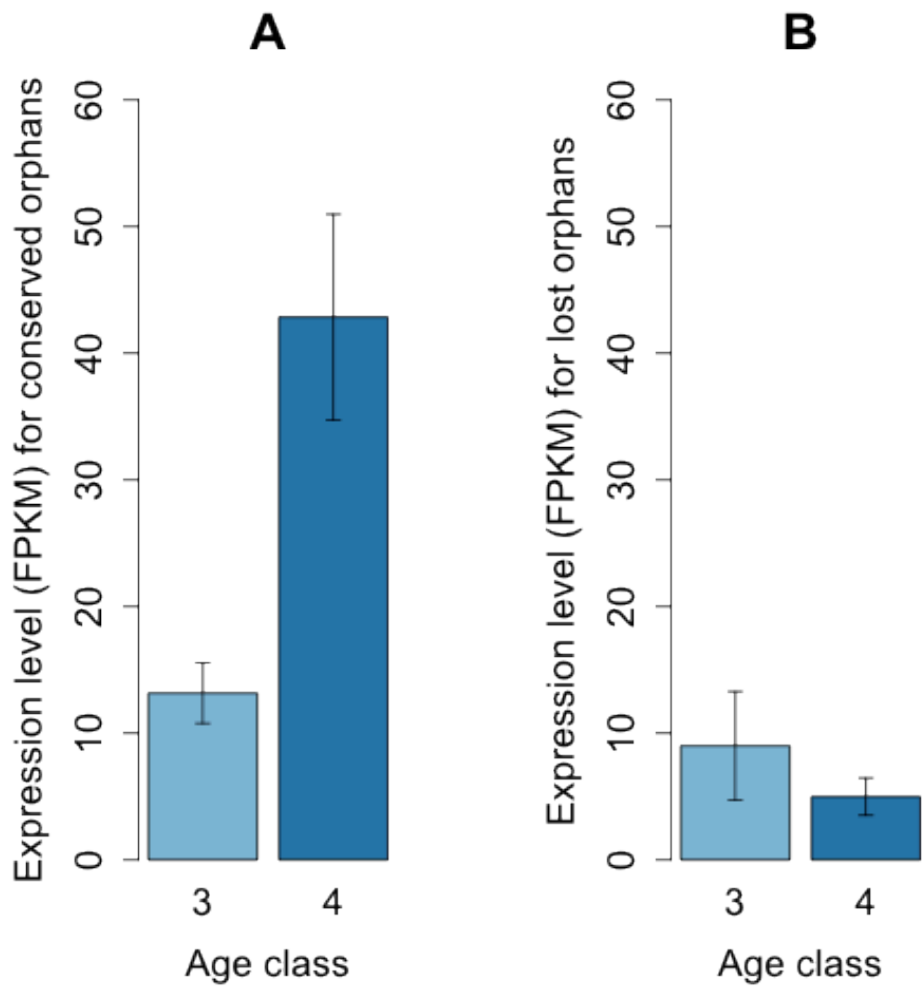

**Figure 16: Conserved and lost orphans differ in their gene expression pattern**

Expression intensity and sex bias in *D. miranda* for orphans conserved in all the *obscura* species (conserved orphans) *vs.* orphans that pseudogenized in *D. lowei* and/or *D. persimilis* (lost orphans). Expression is calculated in males for orphans of age classes 3 and 4. Expression level increases with age for conserved orphans (A), while it decreases for lost orphans (B).

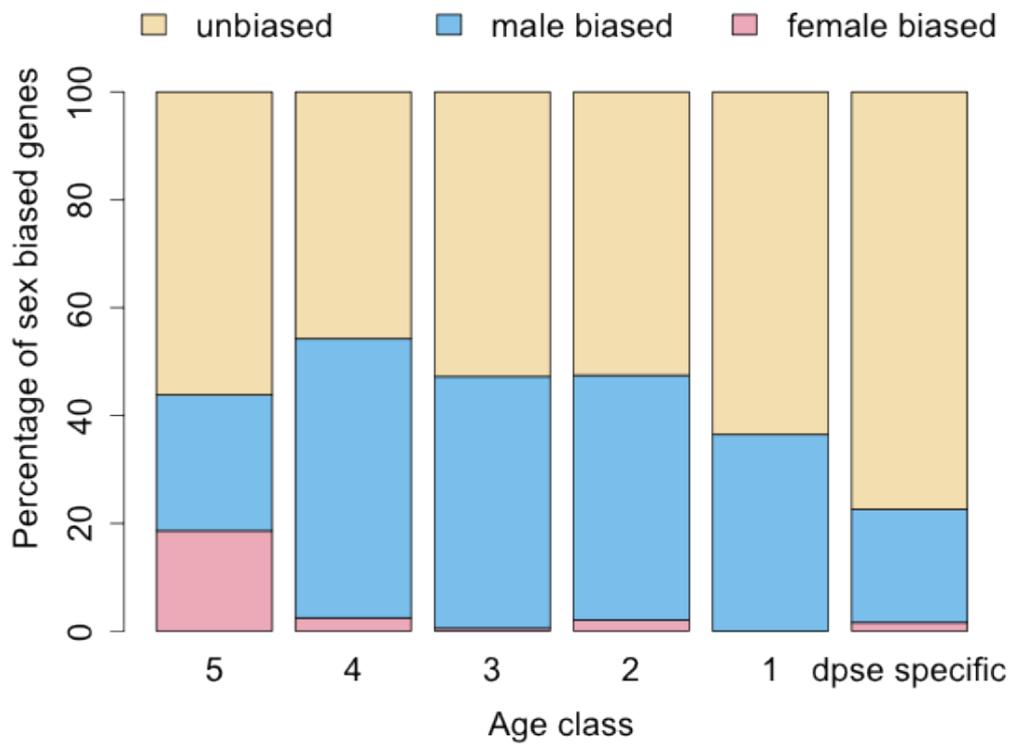

**Figure 17: The proportion of male-biased orphans increases with age**

Sex-biased expression was measured in *D. pseudoobscura* for orphans belonging to different age classes and for old genes (age class 5).

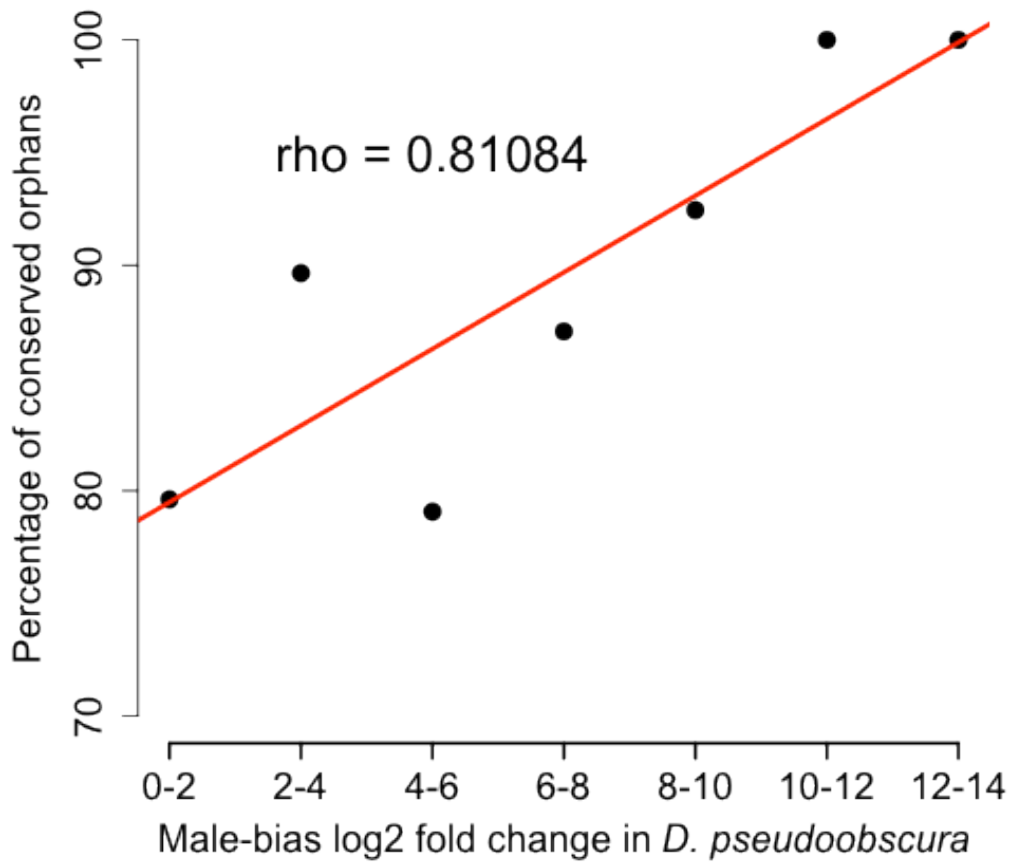

**Figure 18: Conservation of orphans is correlated with male-biased gene expression**

Orphans with male-biased gene expression in *D. pseudoobscura* were grouped into classes according to expression bias strength. The fraction of conserved orphans in each bin shows a significant positive correlation with expression bias (Spearman's *rho* = 0.811, P = 0.02692). This correlation suggests that orphans with a more prounounced male-biased expression tend to persist longer than less male-biased orphans. No similar trend was seen for female-biased orphans (Spearman's *rho* = 0.78262, P = 0.1176).

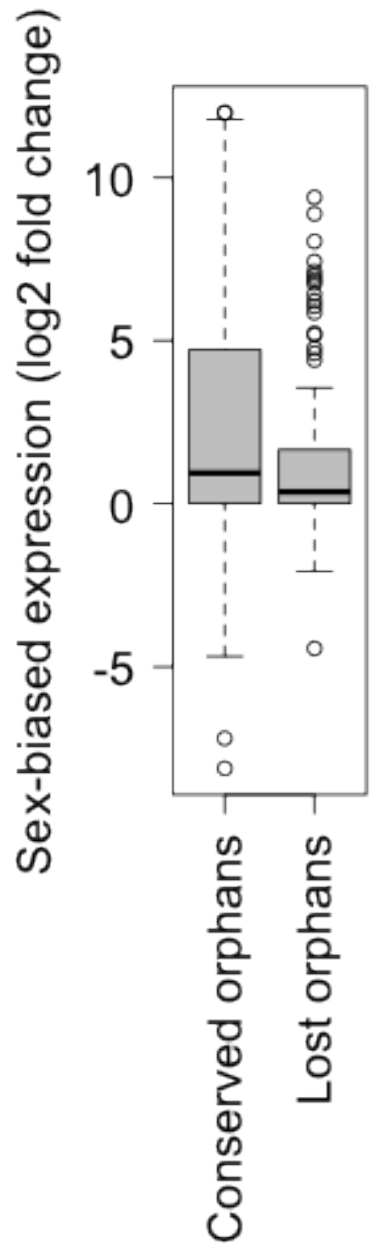

**Figure 19: Comparison of strength of sex-biased gene expression for conserved and lost orphans in *D. miranda***

A sex-biased gene expression larger than zero indicates a higher gene expression intensity in males than in females (male-biased gene expression). Conserved orphans have significantly higher male-biased expression than lost orphans (Mann-Whitney test, P = 0.03158).

# Supplementary files

**Supplementary File 1: Schematic tree of the *Drosophila* species analyzed in this study**

The tree includes the 12 *Drosophila* species from FlyBase (Clark, Eisen et al. 2007) plus three additional members of the obscura group (*D. affinis*, *D. lowei* and *D. miranda*). The *obscura* group is highlghted in magenta. The species corresponding to the black subtrees were used as outgroups in the orphan detection pipeline (see Methods). Divergence times for the 12 *Drosophila* species are taken from Table 3 in (Obbard, Maclennan et al. 2012) (estimates based on mutation rate); for *D. affinis* and *D. miranda* from (Gao, Watabe et al. 2007); for *D. lowei* from (Beckenbach, Wei et al. 1993).

**Supplementary File 2: *De novo* assembly statistics**

**Supplementary File 3: Orthology annotation statistics**

**Supplementary File 4: Summary table of orphans and old genes**

From left to right the columns refer to: gene id, gene name, gene class (orphan/old), age class, chromosome, gene start, gene end, gene strand, length of the longest transcript, corresponding protein length, number of introns, GC content of the coding sequence, Codon Adaptation Index, expression levels in *D. pseudoobscura* males and females in FPKM, expression levels in *D. miranda* males and females normalized using the size factors normalization from the R package DESeq (Anders and Huber 2010), *dN/dS* in *D. miranda / D. pseudoobscura*.